\documentclass[apj,numberedappendix]{emulateapj}

\slugcomment{Accepted for publication in ApJ} 

\usepackage{graphicx}
\usepackage{bm}
\usepackage{CJK}
\usepackage{color}

\usepackage[colorlinks=true,citecolor=blue,linkcolor=magenta,urlcolor=cyan]{hyperref}
\usepackage{ctable}

\shorttitle{Non-radial Eruption}
\shortauthors{Sun et al.}

\begin{document}

\begin{CJK}{UTF8}{}

\title{A Non-radial Eruption in a Quadrupolar\\
Magnetic Configuration With a Coronal Null}

\author{
\begin{CJK}{UTF8}{gbsn} 
Xudong Sun (孙旭东)\altaffilmark{1,2}, J. Todd Hoeksema\altaffilmark{1}, Yang Liu (刘扬)\altaffilmark{1}, \\ Qingrong Chen (陈庆荣)\altaffilmark{2}, \end{CJK} \begin{CJK}{UTF8}{min}Keiji Hayashi (林啓志)\altaffilmark{1}\end{CJK}
}

\email{xudong@sun.stanford.edu}

\altaffiltext{1}{W. W. Hansen Experimental Physics Laboratory, Stanford University, Stanford, CA 94305, USA.}
\altaffiltext{2}{Department of Physics, Stanford University, Stanford, CA 94305, USA.}


\begin{abstract}
We report one of several homologous non-radial eruptions from NOAA active region (AR) 11158 that are strongly modulated by the local magnetic field as observed with the \textit{Solar Dynamic Observatory} (\textit{SDO}). A small bipole emerged in the sunspot complex and subsequently created a quadrupolar flux system. Non-linear force-free field (NLFFF) extrapolation from vector magnetograms reveals its energetic nature: the fast-shearing bipole accumulated $\sim$2$\times$10$^{31}$ erg free energy (10$\%$ of AR total) over just one day despite its relatively small magnetic flux (5$\%$ of AR total). During the eruption, the ejected plasma followed a highly inclined trajectory, over 60$^\circ$ with respect to the radial direction, forming a jet-like, inverted-Y shaped structure in its wake. Field extrapolation suggests complicated magnetic connectivity with a coronal null point, which is favorable of reconnection between different flux components in the quadrupolar system. Indeed, multiple pairs of flare ribbons brightened simultaneously, and coronal reconnection signatures appeared near the inferred null. Part of the magnetic setting resembles that of a blowout-type jet; the observed inverted-Y structure likely outlines the open field lines along the separatrix surface. Owing to the asymmetrical photospheric flux distribution, the confining magnetic pressure decreases much faster horizontally than upward. This special field geometry likely guided the non-radial eruption during its initial stage.
\end{abstract}

\keywords{Sun: activity --- Sun: corona --- Sun: surface magnetism --- Sun: magnetic topology}


\section{Introduction}
\label{sec:intro}

Solar eruptive events derive their energy from the non-potential coronal magnetic field \citep{forbes2000,hudson2011}. Reconnection takes place locally where the field gradient is large, but can alter the larger-scale field topology rapidly. The dissipated energy from the relaxing field accelerates particles, produces radiation, and heats and ejects plasma into the interplanetary space as a coronal mass ejection (CME).

Prior to eruption, energy builds up in the corona through flux emergence and displacement, which may take up to a couple of days \citep{schrijver2009}. The slow evolution can be approximated by a series of quasi-stationary, force-free states in the low plasma-$\beta$ coronal environment. This allows the estimation of AR energetics in non-flaring states, thanks to recent advances in photospheric field measurement and field extrapolation algorithms \citep{regnier2006,thalmann2008,jing2009,sun2012}.

Besides the gross energy budget, the detailed magnetic configuration also proves important to the initiation, geometry, and scale of eruptions. In the case of a coronal jet, the direction of the ambient field (horizontal or oblique) directly determines the direction of the jet and its distinct emission features \citep[two-sided or ``anemone'' type,][]{shibata1997}. Observation and modeling demonstrate that the overlying field provides a critical constraint on CME's speed and trajectory \citep{liuyang2007,gopalswamy2009,wangyuming2011}.

Theoretical studies have extensively explored the role of topological features in reconnection \citep{demoulin1996,priest2000,longcope2005}. Their applications to solar events usually involved the results of potential or linear force-free field extrapolation \citep{aulanier2000,fletcher2001,mandrini2006}, or magnetohydrodynamic (MHD) simulations that qualitatively reproduce the observed phenomena \citep{moreno2008,pariat2009,masson2009,torok2011}.

Here we report one of several similar non-radial eruptions that are strongly modulated by the local magnetic field as observed with the \textit{Solar Dynamic Observatory} (\textit{SDO}). Using vector magnetograms from the Helioseismic and Magnetic Imager (HMI) \citep{schou2012,hoeksema2012} aboard \textit{SDO} and a non-linear force-free field (NLFFF) extrapolation, we monitor the AR evolution and explain the magnetic topology that leads to the curious features during the eruption. The Atmospheric Imaging Assembly \citep[AIA;][]{lemen2012} and other observatories recorded these features and provide guidance for our interpretation.

In Section~\ref{sec:method} we briefly describe the data and the extrapolation algorithm. We first present observations of the eruption in Section~\ref{sec:erupt}, and then come back in Section~\ref{sec:evo} to explain the magnetic field and energy evolution leading to the event. In Section~\ref{sec:topo}, we interpret this curious event based on the magnetic field topology. We discuss the interpretation in Section~\ref{sec:discuss} and summarize in Section~\ref{sec:summary}.


\begin{figure}[t!]
\centerline{\includegraphics[width=3in]{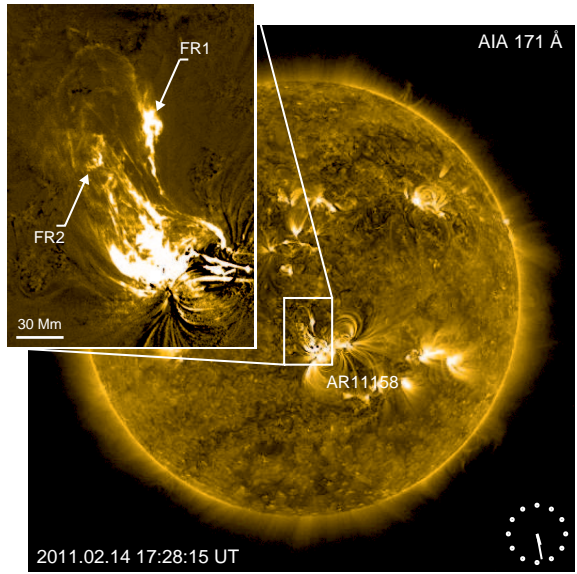}}
\caption{Full-disk, unsharp masked AIA 171 {\AA} image at 17:28:15 UT on February 14, 2011 showing the non-radial eruption. Inset shows the enhanced image of the ejecta. The two flux-rope-like structures with a shared eastern footpoint are marked as FR1 and FR2. Animation of a 20-hr interval shows at least five similar eruptions. (An animation of this figure is available at \url{http://sun.stanford.edu/~xudong/Article/Cusp/homolog.mp4}.) \label{f:homolog}}
\end{figure}

\begin{figure*}[t!]
\centerline{\includegraphics{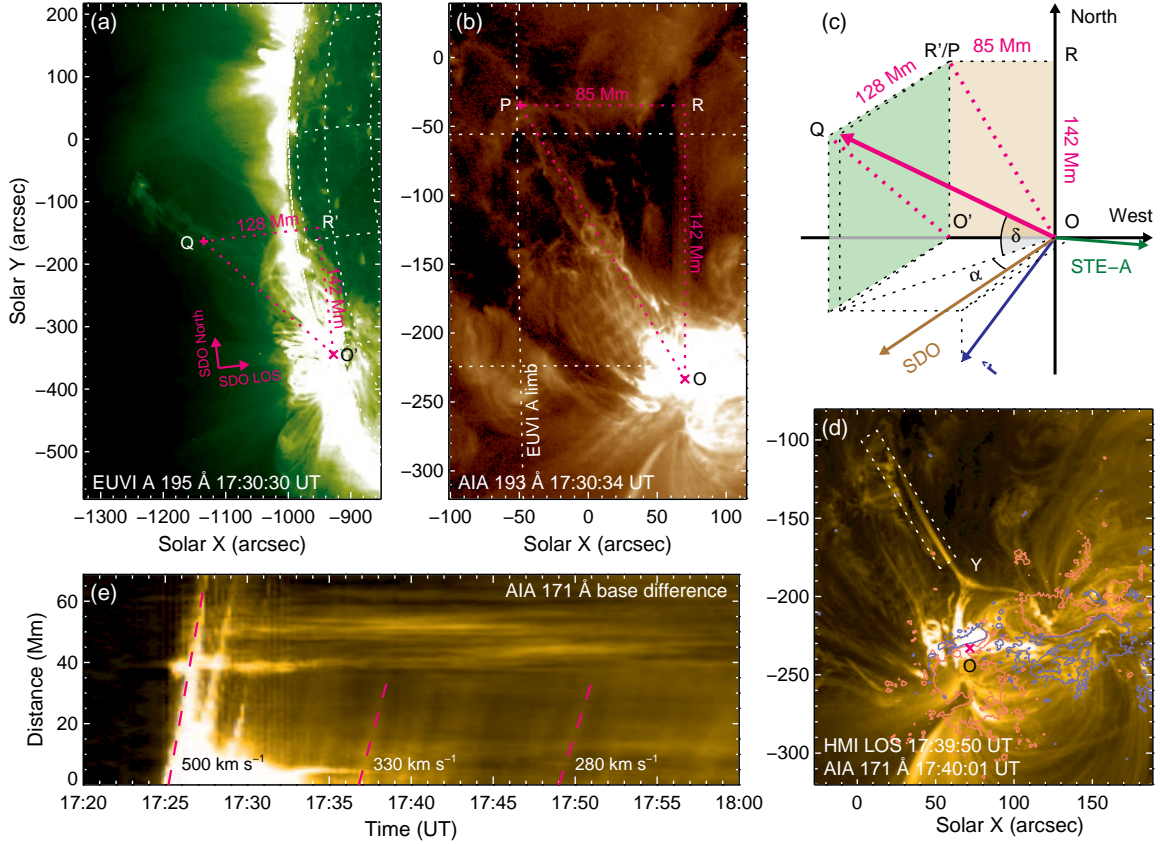}}
\caption{Geometry of the non-radial eruption. O and $\rm{{O'}}$ mark the eruption site. (a) SECCHI EUVI 195 {\AA} image from \textit{STEREO}-A, about 87$^\circ$ ahead of \textit{SDO}. Due to the tilt of the solar rotational axis, the \textit{SDO} and \textit{STEREO} north are offset by 6.8$^\circ$. (b) AIA 193 {\AA} image of the same ejecta, taken 4 s later than (a). The projected N-S length of the ejecta ($|{\rm{OR}}|$) is identical to that in (a) ($|\rm{{O'R'}}|$), where ${\rm{OR}}$ and ${\rm{O'R'}}$ represent the projection of line segments OP and OQ in the N-S direction in \textit{SDO}'s plane-of-sky, respectively. The scales of (a) and (b) are different in order to better show the features of interest. (c) Schematic diagram explaining the determination of the ejecta's geometry. SDO's west, north, and LOS directions are taken as $x$, $y$, and $z$ axis. The pink arrow represents the ejecta, its projected shape viewed from EUVI and AIA are shown as pink dashed lines on green and brown planes. The local radial vector is about W13S04 to LOS. The inclination $\delta$ is about 43$^\circ$; azimuth $\alpha$ about 34$^\circ$. See Section~\ref{sec:erupt} for details. (d) AIA 171 {\AA} image of the post-eruption AR; Y marks the top of the cusp and the base of the jet. The boxed region is used to construct panel (e). Purple/pink contours are for HMI LOS field at $\pm$200 G. (e) Space-time diagram showing the speed of ejecta and jet. Three dashed lines (starting near 17:25, 17:36, and 17:49 UT) indicate a projected speed of 500, 330 and 280 km s$^{-1}$, respectively. Panels (a), (b), and (d) are displayed in a square-root scale. (An animation of this figure is available at \url{http://sun.stanford.edu/~xudong/Article/Cusp/ejecta.mp4}.) \label{f:ejecta}}
\end{figure*}


\section{Data and Modeling}
\label{sec:method}

Sunspot complex AR 11158 produced the first X-class flare of cycle 24 near its center on 2012 February 15 \citep{schrijver2011x}. Before and after that flare, there were a series of smaller eruptions from its northeastern periphery, our region of interest (ROI), where a small new bipole emerged. Five of them assumed very similar structures and were accompanied by C or M-class flares within a 20-hr interval (06:58, 12:47, 17:26, and 19:30 UT on February 14, and 00:38 UT on February 15; see the animation of Figure~\ref{f:homolog} and Figure~\ref{f:energy}(d)). In all cases the ejecta followed a similar, non-radial trajectory towards the northeast.

We focus here on the event around 17:26 UT on February 14 associated with an M-2.2 class flare. The eruption site was near central meridian (W04S20). For context, we study the AR field evolution during a 36-hr interval leading to and shortly afterward the event, from February 13 12:00 UT to February 15 00:00 UT.

The HMI vector magnetograms provide photospheric field measurement at 6173 {\AA} with 0.5$\arcsec$ pixels and 12-minute cadence. Stokes parameters are first derived from filtergrams averaged over a 12-minute interval and then inverted through a Milne-Eddington based algorithm, the Very Fast Inversion of the Stokes Vector \citep[VFISV;][]{borrero2011}. The 180$^\circ$ azimuthal ambiguity in the transverse field is removed using an improved version of the ``minimum energy'' algorithm \citep{metcalf1994,leka2009}. Here, the selected 36-hr dataset includes 181 snapshots of a $\sim$300$\arcsec$$\times$300$\arcsec$ region. For data reduction procedures, we refer to \citet{hoeksema2012} and references therein.

We use an optimization-based NLFFF extrapolation algorithm \citep{wiegelmann2004} and HMI data as the lower boundary to compute the coronal field. The side and upper boundaries are determined from a potential field extrapolation (PF) using the Green's function method \citep{sakurai1989}. The computation domain assumes planar geometry, uses a Cartesian grid (300$\times$300$\times$256) and a 720 km ($\sim$1$\arcsec$) resolution. Before extrapolation, we apply to the data a pre-processing procedure \citep{wiegelmann2006} that iteratively reduces the net torque and Lorentz force so the boundary is more consistent with the force-free assumption. The magnetic free energy is simply the energy difference between the NLFFF and PF. Our previous study on the same region \citep{sun2012} used identical procedures, where we described and evaluated the algorithm in detail.


\section{The Non-Radial Eruption}
\label{sec:erupt}

Observed in the AIA extreme-ultraviolet (EUV) bands, a small AR filament situated above the polarity inversion line (PIL) of a newly emerged bipole started its slow rise around 17 UT (see online animation of Figure~\ref{f:ejecta}). The M-class flare peaked at 17:26 UT in soft X-ray (SXR) flux, when the filament rapidly erupted towards the northeast. The ejecta appeared to consist of two rope-like features (FR1 and FR2 in Figure~\ref{f:homolog}) with a shared eastern footpoint. By inspecting AIA image sequences in various bands and HMI magnetograms, we think that they originated from the same filament structure.

The \textit{STEREO}-A spacecraft was then near quadrature with the Sun-Earth line (87$^\circ$ ahead). Its SECCHI EUVI instrument \citep{howard2008} caught a glimpse of the ejecta in the 195 {\AA} channel (Figures~\ref{f:ejecta}(a)), where the erupted filament appeared to follow a straight trajectory viewed from west. Using the simultaneous image from the AIA 193 {\AA} channel (Figures~\ref{f:ejecta}(b)), we are able to estimate its three-dimensional (3D) geometry.

Figure~\ref{f:ejecta}(c) illustrates the triangulation procedure. We manually select the eruption site O and the frontmost point P of the inner flux rope FR2 (as projected on the plane of sky) in the AIA image. We select the corresponding points ${\rm{O'}}$ and Q in the EUVI image, such that 1) O and ${\rm{O'}}$ have the same Carrington coordinate; 2) the ejecta's N-S extent in two images satisfies ${\rm{|OR|}}={\rm{|O'R'|}}$, where ${\rm{OR}}$ and ${\rm{O'R'}}$ represent the projection of line segments OP and OQ in the N-S direction in \textit{SDO}'s plane-of-sky, respectively. 

Assuming the ejecta follows a straight trajectory, we can solve for its inclination $\delta$ and azimuth $\alpha$ with respect to the line-of-sight (LOS). We find that $\delta$=43$^\circ$, $\alpha$=34$^\circ$. By repeating the point selection process we estimate the uncertainty to be $\sim$3$^\circ$ under the current scheme. The trajectory is highly inclined, about $66^\circ$ with respect to the local radial direction.

\begin{figure*}[t!]
\centerline{\includegraphics{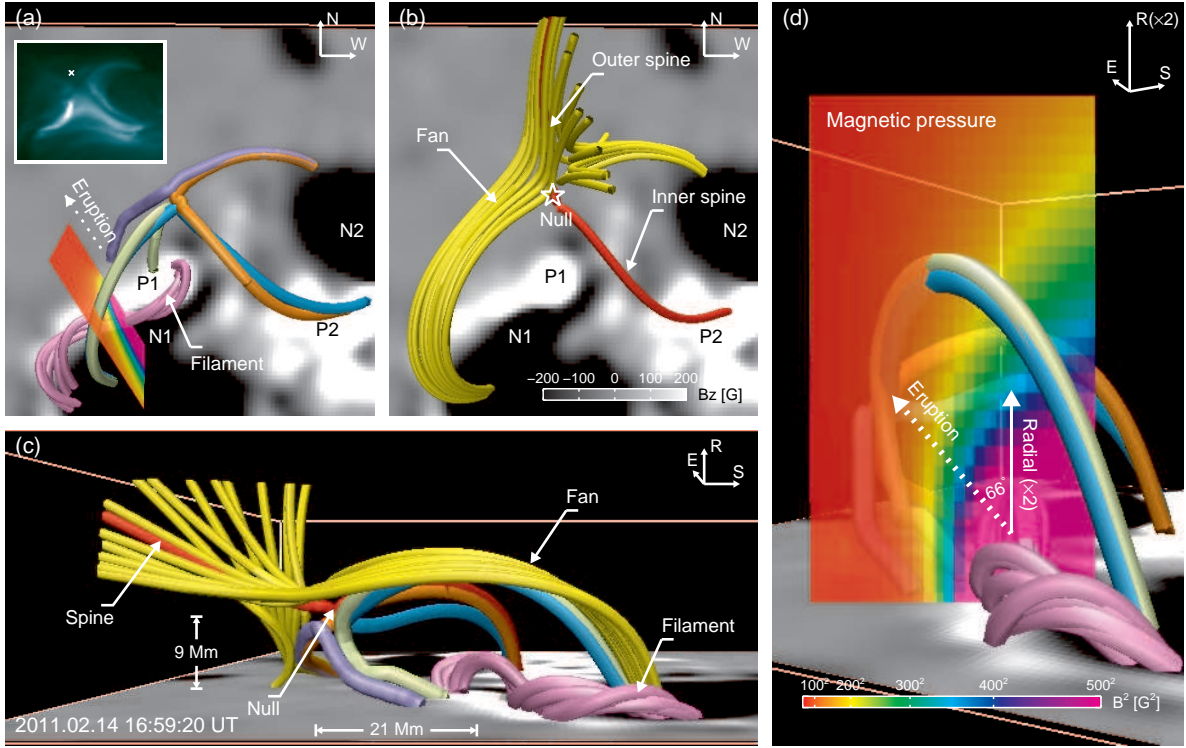}}
\caption{Magnetic topology based on NLFFF extrapolation for the pre-eruption state. (a) SDO view of four sets of loops connecting the four quadrupolar flux components pairwise, as well as twisted field lines below representing the AR filament. The cross section is identical to that in (d). Inset shows the corresponding AIA 94 {\AA} image, which is the same as Figure~\ref{f:corona}(a). The inferred coronal null point, marked by ``X'', appears slightly above the observed loops. (b) Magnetic null point, spine field line, and open field lines that outline the separatrix (fan) surface. (c) Side view of the region (from east). (d) Side view with $z$-axis (radial direction) stretched by 2. Magnetic pressure is imaged on a vertical cross section to illustrate its anisotropy. The cross section is roughly aligned with the direction of eruption, and is in front of the null from this viewing angle. (An animation of this figure is available at \url{http://sun.stanford.edu/~xudong/Article/Cusp/topo.mp4}.) \label{f:topo}}
\end{figure*}

A bright, inverted-Y shaped structure formed in the wake of the eruption. It consisted of a thin spire on top of a cusp-shaped loop (Figure~\ref{f:ejecta}(d)); both lasted over 1 hr. The cusp appeared almost two-dimensional and had both ``legs'' rooted in negative polarity flux (see Section~\ref{sec:topo} and Figure~\ref{f:topo}(b)). There were propagating brightness disturbances along the cusp legs and the spire \citep[][see the animation of Figure~\ref{f:ejecta}]{thompson2011}, which have been interpreted as episodic plasma flows \citep[see the coronal seismic and Doppler analyses in][]{sujt2012,tian2012}. These observed features outline a magnetic arrangement that resembles a coronal jet \citep[e.g.][]{shibata1997}. Nevertheless, the structure appeared only \textit{after the eruption}. Various observed features appear to require alternative explanations other than the standard jetting model or its variations (see a brief discussion in Section~\ref{subsec:d_mech}).

By placing a cut along the thin spire in the AIA 171 {\AA} image sequence, we construct a space-time diagram to illustrate the relevant speeds in this event (Figure~\ref{f:ejecta}(d) and (e)). The projected speed of the ejecta is about 500 km s$^{-1}$; the brightness disturbance is around 300 km s$^{-1}$. Considering the inclined trajectory, we estimate the real speed about 30$\%$ higher, i.e. 650 and 390 km s$^{-1}$, respectively.


\begin{figure*}[t!]
\centerline{\includegraphics[width=5.2in]{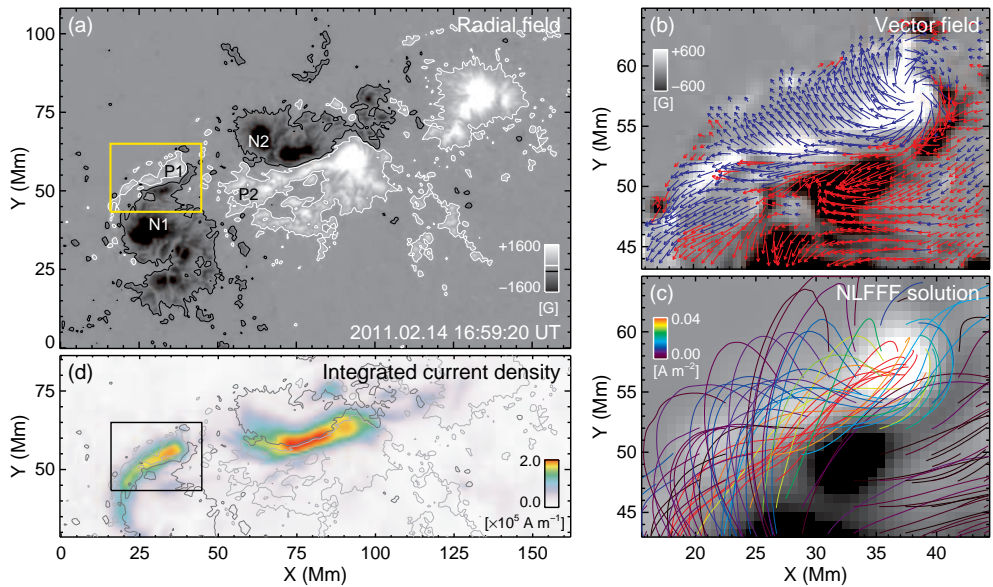}}
\caption{Snapshot of magnetic field of AR 11158, 25 minutes before the eruption. (a) Radial magnetic field ($B_r$) map as derived from the vector magnetogram. The contours are for $\pm$200 G. P1, N1, P2, and N2 mark four components of the quadrupolar flux system. The yellow box indicates the FOV for (b) and (c) and is identical to that in (d). (b) Photospheric vector magnetic field map. Gray-scale background shows $B_r$. The blue/red arrows indicate the horizontal component (${\bf{B}}_h$) with positive/negative radial counterpart, where field strength $B>200$ G. Their lengths correspond to the magnitude ($B_h$); their directions show the azimuth. (c) Selective extrapolated field lines plotted on $B_z$ map. The color shows the amount of radial current at the field line footpoint. (d) Map of current density ($|J|$) integrated over the lowest 10 Mm in extrapolated field. The light/dark gray contours are for $B_r=\pm$200 G. All data are deprojected and remapped using the Lambert equal area projection. (An animation of this figure is available at \url{http://sun.stanford.edu/~xudong/Article/Cusp/field.mp4}.) \label{f:field}}
\end{figure*}

\begin{figure}[b!]
\vspace{8pt}
\centerline{\includegraphics[width=3.0in]{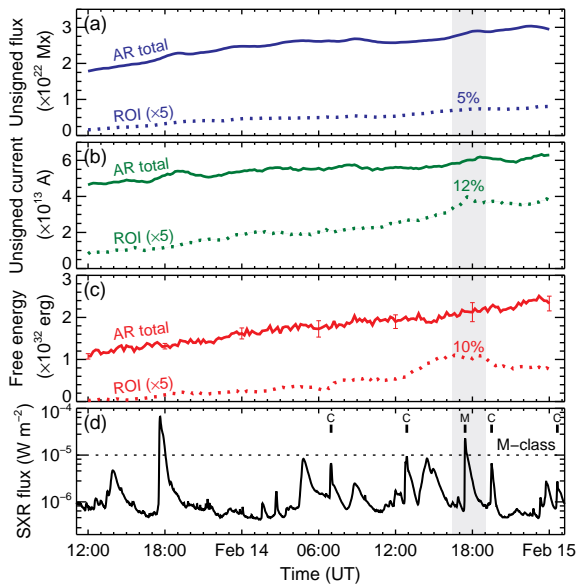}}
\caption{Evolution of the entire AR and the region of interest (ROI). (a) Unsigned magnetic flux. (b) Unsigned photospheric electric current. (c) Magnetic free energy evaluated in the volume above the ROI. (d) \textit{GOES} soft X-ray flux. The homologous flares are marked with ``M'' and ``C'' according to their $GOES$ X-ray class. In panels (a)-(c), solid line represents the whole AR. Dotted line is for the bipole (boxed region in Figure~\ref{f:field}(a)); bipole values are multiplied by 5 for clarity. The gray vertical band indicates a 2.6-hr period that brackets the eruption, whose means are compared with the AR means. For (a) (b) only pixels with $B>200$ G are included. Errors in (a) (b) derived from spectropolarimetry inversion are small. Errors (3$\sigma$) in panel (c) are evaluated using a pseudo Monte-Carlo method \citep{sun2012} which show the effect of spectropolarimetric noise. \label{f:energy}}
\end{figure}


\section{The Emerging Bipole As Energy Source}
\label{sec:evo}

We study the underlying photospheric field that led to this eruption. Figure~\ref{f:field}(a) shows a snapshot of the radial field taken 25 minutes before the event as derived from the vector magnetogram. The AR mainly consists of two interacting bipoles. A large amount of magnetic free energy was stored near the major PIL between the shearing sunspots at center of the field of view (FOV), where the X-class flare took place \citep{sun2012}.

The eruption studied here is related to a newly emerged, smaller bipole (boxed region in Figure~\ref{f:field}(a)). The bipole appeared on February 13 in the northeastern part of the AR. Starting from 12 UT on February 14, the positive component advanced rapidly westward with strong rotational motion and shearing with respect to its negative counterpart, leaving behind a fragmented stripe of flux mimicking a long-tailed tadpole (see the online animation of Figure~\ref{f:field}).

The new bipole had strong horizontal photospheric field that that lay parallel to the PIL (Figure~\ref{f:field}(b)). NLFFF extrapolation suggests a highly twisted core field and strong radial current (Figure~\ref{f:field}(c)), which correspond to the observed AR filament that eventually erupted.

We summarize in Figure~\ref{f:energy} the bipole's temporal evolution. By evaluating the area within the ROI (boxed region in Figure~\ref{f:field}(a)), we estimate its unsigned flux to be only about 5$\%$ of the AR's total around the eruption time. However, the surface unsigned radial current with in the ROI accounts for 12$\%$ of the AR's total, much higher than the corresponding flux fraction. We integrate the free energy in the volume above the ROI and find it to be over 10$\%$ of that in the whole volume. For the ROI, the ratio between the NLFFF energy and the PF energy is about 1.60. This indicates the bipole is very non-potential and energetic. There is a strong concentration of current near the PIL in the lower corona, similar to the major PIL near the center of the AR (Figure~\ref{f:field}(d)).

Unfortunately, we do not find a clear, step-wise change in free energy during the flare that can be used as a proxy of the energy budget (Figure~\ref{f:energy}(c)). Our earlier work on the ensuing X-class flare \citep{sun2012} suggests the energy budget tends to be underestimated by the extrapolation method. This is partly because the flaring field is dynamic and likely not force-free \citep[e.g.][]{gary2001}; thus it cannot be reliably described by the NLFFF model. Limited resolution and uncertainties in the field measurement and modeling may also be a factor. The free energy for the ROI gradually decreased after 20 UT when the positive flux fragmented and the current decreased.

The emergence of the bipole led to a local enhancement of free energy with a series of ensuing eruptions from this relatively small region. Its very existence changed the original magnetic configuration and converted it into an asymmetrical (the new bipole is relatively small) quadrupolar flux system. The change of the photospheric flux distribution altered the coronal magnetic connectivity in a fundamental way, and may have contributed to the destabilization of the system. For clarity, we label the four quadrupolar components P1, N1 (including the old sunspot and the negative part of the new bipole), P2, and N2 (Figure~\ref{f:field}(a)).


\begin{figure*}
\centerline{\includegraphics{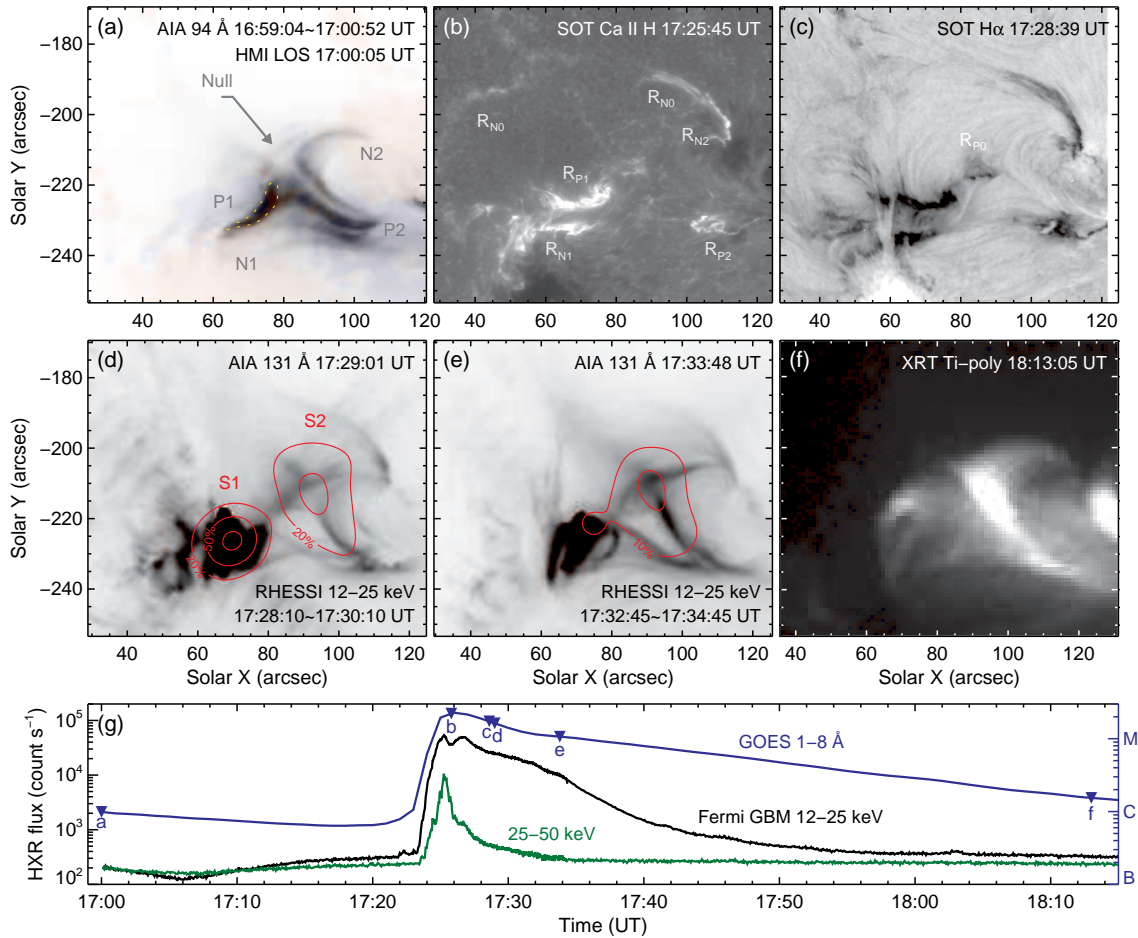}}
\caption{Various flare emission observations. The time of each image is marked in (g). (a) Composite of negative AIA 94 {\AA} image ($\sim$6 MK, 2-minute average) and HMI LOS magnetogram showing the possible coronal null point. The AR filament is manually outlined with dotted line. The null is at (77$\arcsec$, -209$\arcsec$) on the plane of sky. See also Figure~\ref{f:topo}(a). (b) \textit{Hinode}/SOT Ca II H band showing flare ribbons (${\rm{R_{P1}}}$, ${\rm{R_{P2}}}$, ${\rm{R_{N1}}}$, ${\rm{R_{N2}}}$) and the underlying photosphere. Half-ring-like secondary ribbon (${\rm{R_{N0}}}$) is also visible. (c) Unsharp masked, negative SOT H$\alpha$ image showing flare ribbons and magnetic connectivity. An additional, weak brightening is marked as ${\rm{R_{P0}}}$. The erupting plasma is visible in the foreground from about (60$\arcsec$, -250$\arcsec$) upward; the tip of the cusp is near (45$\arcsec$, -160$\arcsec$). (d) \textit{RHESSI} 12-25 keV HXR image as contours on negative AIA 131 {\AA} image ($\sim$10 MK), showing a footpoint source (S1) and coronal source (S2). The two ribbons ${\rm{R_{P1}}}$ and ${\rm{R_{N1}}}$ are spatially unresolved in HXR. Contours are drawn at 20$\%$, 50$\%$, and 90$\%$ of maximum. (e) Same as (d), for 4 minutes later. Contours are for 10$\%$ and25$\%$ of the maximum of (d). (f) \textit{Hinode}/XRT SXR image, showing the cusp-like structure. (g) \textit{GOES} SXR flux, \textit{Fermi}/GBM 12-15 keV, and 25-50 keV HXR flux. \textit{RHESSI} coverage of the flare started from 17:27:44 UT, and is not shown here. Panels (b)-(e) are displayed using square-root scale. All images are tracked with solar rotation and co-aligned to an accuracy better than 0.6{\arcsec}. \label{f:corona}}
\end{figure*}


\section{Interpretation Based on the Magnetic Field Topology}
\label{sec:topo}

\subsection{A Coronal Null and the Inclined Trajectory}
\label{subsec:null}

What is the coronal magnetic field topology that led to the highly non-radial eruption? Field lines computed from the pre-flare NLFFF solution (16:59 UT) reveal connectivity between each pair of the opposite polarity flux (P1/N1, P2/N1, P2/N2, and P1/N2) in this quadrupolar system (Figure~\ref{f:topo}(a)). Such connectivity is apparent in the AIA observations.

One striking feature, however, is the large gradient in field line mapping. For example, loops connecting P2/N1 (cyan) and P2/N2 (orange) are at first parallel, but diverge drastically near their apexes, becoming almost antiparallel with each other. These modeled field lines closely resemble the observed loops (inset of Figure~\ref{f:topo}(a) or Figure~\ref{f:corona}(a)). The cusp-like P2/N2 and the diverging field lines strongly suggest the existence of a coronal null point, where field strength becomes zero.\footnote{See \textit{TRACE} observation of AR 9147/9149 (\url{http://trace.lmsal.com/POD/TRACEpodarchive4.html}).}

Using a trilinear method \citep{haynes2007}, we indeed find a null point situated at $\sim$9 Mm height (Figure~\ref{f:topo}(b)) right above the modeled loop apexes (see Appendix~\ref{a:skeleton}). From that null, closed loops ``turn away'' with a sharp angle. Seen from side (Figure~\ref{f:topo}(c)), these loops are low-lying; they incline towards the northeast, the direction of the eruption. This configuration persisted over the next few hours (see Section~\ref{subsec:d_topo}).

This inclined geometry is perhaps a natural consequence of the asymmetrical photospheric flux distribution. We infer that this field configuration may have facilitated the non-radial eruption in the following ways. First, reconnection may take place near the null point, removing the overlying flux above P1/N1 and preferentially reducing the confinement from the northeast direction. Second, the ambient, confining magnetic pressure ($p_B$=$\frac{B^2}{8\pi}$) is anisotropic: it drops off much faster horizontally than it does in the radial direction (Figure~\ref{f:topo}(d)). When the anisotropy is strong enough, it can guide the ejecta towards a direction with large negative pressure gradient by deflecting its trajectory. It effectively creates a non-radial ``channel'' for the plasma to escape.

\subsection{The Inverted-Y Structure}
\label{subsec:jet}

We further analyze the magnetic topology of the pre-eruption state for insight on the observed inverted-Y shaped structure. By analyzing the Jacobian field matrix (${\rm{M}}_{ij}=\partial B_i / \partial x_j$) at the inferred null point, we are able to find the spine and the fan, which are special field lines that define the magnetic configuration near the singularity \citep[e.g.][]{parnell1996}. Regular field lines passing by the immediate vicinity of the null point generally outline the separatrix (fan) surface (Figure~\ref{f:topo}(b)(c)). In this case they separate the closed flux inside and the open flux outside. We describe the analysis method in Appendix~\ref{a:skeleton}.

Owing to the local excess of negative flux, open field lines from N1 and N2 flow along the separatrix and converge around the outer spine. These field lines naturally form an inverted-Y structure (Figure~\ref{f:topo}(b)). Its morphology resembles the observed loops, although less inclined towards the northeast. Their detailed geometry took shape during the the dynamic eruption, which the static extrapolation is unable to model.

\subsection{Observational Evidences}
\label{subsec:evidence}

Because field line mapping diverges and links the whole quadrupolar system, we expect electrons accelerated during the flare near the null point to precipitate along different loop paths, resulting in multiple pairs of flare ribbons brighting simultaneously \citep{shibata1995}. Taken by the Solar Optical Telescope \citep[SOT;][]{tsuneta2008} on the \textit{Hinode} satellite, Ca II H band images (Figure~\ref{f:corona}(b)) indeed show such phenomena. The typical double ribbons (${\rm{R_{P1}}}$/${\rm{R_{N1}}}$) are related to the erupting filament, whereas ${\rm{R_{P2}}}$ and ${\rm{R_{N2}}}$ are likely related to the reconnecting P2/N2 loop. H$\alpha$ images (Figure~\ref{f:corona}(c)) provide additional information on the magnetic connectivity between ${\rm{R_{P1}}}$/${\rm{R_{N2}}}$ and ${\rm{R_{P2}}}$/${\rm{R_{N1}}}$. Remarkably, the ribbon $\rm{R_{P2}}$ appears to be co-spatial with the inferred spine field line footpoint (Figure~\ref{f:topo}(b)), which moved with time as seen in the Ca II H and H$\alpha$ image sequences.

The \textit{Ramaty High Energy Solar Spectroscopic Imager} \citep[\textit{RHESSI};][]{lin2002} missed the impulsive phase but captured what appeared to be a coronal hard X-ray (HXR) source (S2 in Figure~\ref{f:corona}(d)) in the flare's early decaying phase. The source's proximity to the inferred coronal null gives strong support to our interpretation. From the loop top, energetic electrons followed very inclined paths towards the footpoints in P1/N1, which created the footpoint source (S1) corresponding to the ${\rm{R_{P1}}}$/${\rm{R_{N1}}}$ ribbons. This coronal source lasted well into the decaying phase (Figure~\ref{f:corona}(e)).


\section{Discussion}
\label{sec:discuss}

\subsection{On the Coronal Field Topology}
\label{subsec:d_topo}

How common is the magnetic topology determined here? A previous study focused on the quadrupolar configuration of AR 10486 during the 2003 X-17 flare \citep{mandrini2006}. The major eruption was found to involve reconnection at the quasi-separatrix layers \citep[QSL;][]{demoulin1996}, while a smaller brightening was associated with a similar coronal null point determined using a linear force-free extrapolation. In another quadrupolar region AR 11183, similar cusp and jet structures existed at a much larger scale \citep{filippov2012}. The white-light jet extended over multiple solar radii.

We analyze the entire 36-hr series, searching for consistency in time. The coronal null at 9 Mm appeared in a few frames early on February 14, distinct from all other candidates which were mostly below 4 Mm in weak field regions. Starting from 15:35 UT, it appeared at a nearly constant location (within 3 Mm of the first detected null) in over half the frames afterwards (22/42, until February 15 00:00 UT), while the near-surface nulls rarely repeated in two consecutive time steps. We have applied a different null-searching method based on the Poincar\'{e} index theorem \citep{greene1992} and found similar results (23/42, 20 identical to the trilinear method, with 3 additional and 2 missed detections). The repeated detection of null points and the observed homologous eruptions (Figure~\ref{f:energy}(d)) suggest the aforementioned topology is characteristic for this quadrupolar system.

We compute at 1-hr cadence the ``squashing factor'' $Q$ that describes the field mapping gradient \citep{titov2002} by tracing individual field lines and measuring the differences between the two footpoint locations. High-$Q$ isosurface corresponds to QSLs. By inspecting the contour of $Q$ at different heights, we find that multiple QSL's tend to converge and intersect at about 9 Mm. Near the intersection, the field strength is weak, and the field line mapping gradient is invariably large, with or without null point. This illustrates the robustness of our interpretation despite the uncertainties in the extrapolation algorithm \citep[e.g.][]{derosa2009} and the field measurement. (The uncertainties nevertheless can indeed affect the detailed fan-spine configuration, as discussed in Appendix~\ref{a:skeleton}.)

We note that our PF extrapolation, with radial field as boundary condition and the Green's function method, does not detect any nulls above 5 Mm. Instead, we find a low-lying null at about 4 Mm in 13 frames, southwest to the NLFFF solution. The field configuration is less realistic, presumably because the current-free assumption does not agree with observation.

\subsection{On the Flare Emissions}
\label{subsec:d_obs}

Owing to the LOS projection, the altitude of an on-disk HXR source cannot be unambiguously determined. We think S2 is a coronal source mainly because it appeared near the apex of cusp-shaped loops (P2/N2) which is typical for reconnecting field lines \citep[e.g.][]{tsuneta1996}. In addition, its strong HXR emission (peak at $\sim$60$\%$ of the maximum) does not correspond to any bright flare ribbon. The closest chromospheric emission enhancement is a small patch (${\rm{R_{P0}}}$ in Figure~\ref{f:corona}(c)) within a fragmented positive flux about 5$\arcsec$ to the east and south, whose intensity is much weaker than the ${\rm{R_{P1}}}$/${\rm{R_{N1}}}$ ribbons. This argues against the footpoint source interpretation.

We notice a dimmer, half-ring-like ribbon (${\rm{R_{N0}}}$) farther north in the weak field area (Figure~\ref{f:corona}(b)); both H$\alpha$ (Figure~\ref{f:corona}(c)) and EUV images (animation of Figure~\ref{f:ejecta}) show its connection to P1. This structure is related to flux emerging into an encircling unipolar region \citep[``anemone'' AR;][]{shibata1994}. Because the brightening ${\rm{R_{N0}}}$ region possesses flux only a few percent of P1 \citep[c.f.][]{reardon2011}, we consider this structure secondary. It does not affect our conclusions on the AR topology. 

Because no HXR source was detected at the P2/N2 footpoints and the ${\rm{R_{P2}}}$/${\rm{R_{N2}}}$ ribbons were fainter than ${\rm{R_{P1}}}$/${\rm{R_{N1}}}$, we think the electrons primarily precipitated along the shorter P1/N1 loop during the flare. On the other hand, the P2/N2 loop produced much stronger SXR and EUV emission during the flare's late decaying phase. Almost 30 minutes later, SXR images (Figure~\ref{f:corona}(f)) from the \textit{Hinode} X-Ray Telescope \citep[XRT;][]{golub2007} still showed a bright cusp structure above P2/N2.


\begin{figure}[t!]
\centerline{\includegraphics[width=2.6in]{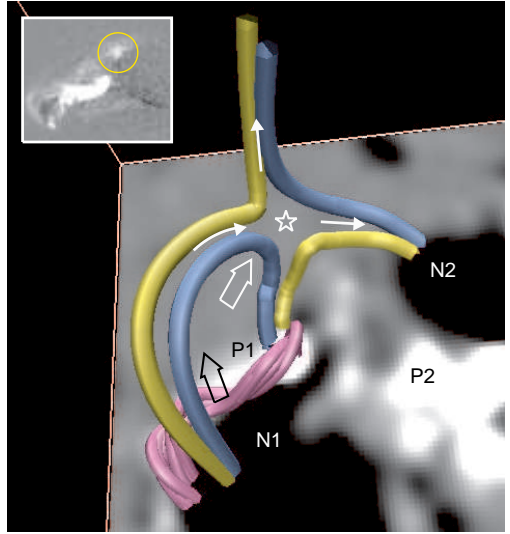}}
\caption{Schematic illustration of the magnetic configuration and dynamics that may have led to the eruption. The structure resembles that of a blowout jet. The arcade (blue field lines above P1/N1) from the newly emerged bipole expands, reconnects with the pre-existing field (blue field lines from N2), becomes open (yellow field lines from N1), and the low-lying sheared/twisted core field (pink field lines between P1/N1) subsequently erupts. A possible initial reconnection site near is marked by the star; possible motions of the loops are denoted by thick arrows. Pre- and post- reconnection field lines are colored blue and yellow, respectively. The directions of the observed, \textit{post-eruption} flow (Figure~\ref{f:ejecta} and animation, see also \cite{thompson2011,sujt2012}) are denoted by thin arrows. The inset shows the SXR difference image between 17:22:32 and 17:19:56 UT from \textit{Hinode} XRT Ti Poly filter (FOV 72{\arcsec}$\times$60{\arcsec}). The brightening P1/N2 loop is marked by a yellow circle; the brightening filament is visible in the foreground. \label{f:jet}}
\end{figure}


\subsection{On the Eruption Mechanism}
\label{subsec:d_mech}

When a bipole emerges, one leg of the new loop may reconnect with the oppositely directed, pre-existing open field. The released magnetic energy heats the plasma and produces field-collimated outward flows, known as the ``standard'' jet phenomenon \citep{shibata1997}. When the emerging field is sheared or twisted, its core may subsequently erupt. Events in this sub-class have recently been described as ``blowout'' jets \citep{moore2010}.

Can this event be explained by the jet models? We find the inferred magnetic structure here resembles the blowout type. Illustrated in Figure~\ref{f:jet}, the newly emerged bipole (P1 and the north part of N1) hosts a twisted core field. We speculate that the increasing flux leads to the expansion of the arcade loops above, which reconnect with the open, negative-polarity field from N2. This process opens up the arcade loops and acts to promote the eventual eruption of the core field below. The jet model predicts the brightening of the reconnected P1/N2 loop, which is indeed observed in the SXR images (inset of Figure~\ref{f:jet}). However, in contrast to the expected jet behavior, no outward flows are observed during this stage. The jet-like, inverted-Y structure appeared only \textit{after} the core field eruption and the accompanying M-class flare.

Propagating brightness disturbances in the post-eruption inverted-Y structure have been interpreted as pulsed plasma flow \citep{sujt2012,tian2012}. The upflow from the left leg diverges and flows in opposite directions, upward in the thin spire and downward in the right leg (Figure~\ref{f:jet} and Animation of Figure~\ref{f:ejecta}). The flow is most pronounced in cooler EUV wavebands (e.g. 171 {\AA}, $\sim$0.6 MK) and is absent in SXR images.  In the standard jet model, these collimated flows are produced and heated by reconnection. The relatively low temperature observed here suggests a low-altitude reconnection site with cooler plasma supply \citep[c.f.][]{sujt2012}, rather than the one near the base of the spire higher in the corona. The detailed dynamics of this event require further investigation which is out of the scope of this work.

\section{Summary}
\label{sec:summary}

We summarize our findings as follows.

\begin{itemize}
\renewcommand{\labelitemi}{$-$}

\item Bipole emergence and shearing in a pre-existing sunspot complex introduced a large amount of free energy, despite its small flux. The new flux powered a series of homologous, non-radial eruptions.

\item One typical eruption had an inclined trajectory about 66$^\circ$ with respect to the radial direction. An inverted-Y structure consisted of cusp and jet formed in the wake of the eruption.

\item The bipole emergence created an asymmetrical quadrupolar flux system. Field extrapolation suggests that the consequent, inclined overlying loops and the anisotropic magnetic pressure are responsible for the non-radial eruption.

\item Extrapolation suggests a coronal null point at about 9 Mm, slightly below the apexes of the cusp-like loops. Its location is favorable for reconnection between different flux components in the quadrupolar system. The observed inverted-Y structure is likely related to the open negative field lines in part outlining the separatrix surface.

\item Multiple flare ribbons brightened simultaneously during the accompanying flare. A coronal HXR source appeared near the inferred null point. These observations support our interpretation.

\item The inferred magnetic structure resembles that of a blowout-type jet. Some observed features fit in the jet model, while others remain difficult to explain.

\end{itemize}

The event studied here demonstrates the importance of detailed magnetic field topology during solar eruptions. Flux emergence in suitable environment can lead to fundamental changes in the coronal field geometry, which then place strong constraints on the plasma dynamics.


\acknowledgments
We thank B. J. Thompson for bringing this event to our attention and the anonymous referee for the helpful comments. We are grateful to T. Wiegelmann for providing the NLFFF extrapolation code. We benefited from discussions with M. Derosa, W. Liu, C.-L. Shen, and L. Tarr. The \textit{SDO} data are courtesy of NASA and the HMI and AIA science teams. We acknowledge the use of \textit{STEREO}/SECCHI EUVI, \textit{Hinode}/SOT, XRT, \textit{RHESSI}, \textit{GOES} and \textit{Fermi}/GBM data. Figures~\ref{f:topo} and \ref{f:jet} are produced by VAPOR (\url{www.vapor.ucar.edu}).

{\it Facilities:} \facility{\textit{SDO}}, \facility{\textit{Hinode}}, \facility{\textit{STEREO}}, \facility{\textit{RHESSI}}, \facility{\textit{GOES}}, \facility{\textit{Fermi}}.



\appendix

\section{Method for Finding The Magnetic Topological Skeleton}
\label{a:skeleton}

At a magnetic null point, field strength becomes 0, and singularity arises. We follow the null-searching method described in \citet{haynes2007}. Assuming the field is trilinear within each volume element, the 3D field vector ${\bf{B}}=(B_1,B_2,B_3)^{\rm{T}}$ and its derivatives ($\partial B_i / \partial x_j$, $i,j=1,2,3$) within each cell are completely determined by the values on its eight vertices. To search for possible null point, we first scan over each cell in the domain: if any $B_i$'s have the same sign on all the vertices, the cell cannot host a null point and will be ignored. For each remaining cell, we use a Newton-Raphson scheme to iteratively solve for ${\bf{x}} = (x_1,x_2,x_3)^{\rm{T}}$ that satisfies $B_i({\bf{x}})=0$:
\begin{equation}
{\bf{x}}^{n+1} = {\bf{x}}^n - \left( \partial{\bf{B}}({\bf{x}}^n) \over \partial{\bf{x}}^n \right)^{-1} {\bf{B}}({\bf{x}}^n),
\end{equation}
where $n$ and $n+1$ denote two consecutive iteration steps, and the repeated index $j$ means summing of all $j$'s. For the 16:59 UT frame, we find a null point at ${\bf{x}} = (89.2425, 173.8625, 12.7667)^{\rm{T}}$ in the (300$\times$300$\times$256) domain. At a 720 km resolution, its height is about 9.2 Mm. The field strength $|B|$ is about $10^{-5}$ G.

The rest of the method description is adapted from \citet{parnell1996} and \citet{haynes2010}. To first order, the magnetic field near a null point located at ${\bf{x}}'$ is approximated by
\begin{equation}
B_i = M_{ij} (x_j-x'_j),
\end{equation}
where the matrix $M_{ij}=\partial B_i / \partial x_j$ is the Jacobian matrix, and is evaluated in this case as
\begin{equation}
M_{ij} = \left( \begin{array}{rrr}
\partial B_1 / \partial x_1 & \partial B_1 / \partial x_2 & \partial B_1 / \partial x_3 \\
\partial B_2 / \partial x_1 & \partial B_2 / \partial x_2 & \partial B_2 / \partial x_3 \\
\partial B_3 / \partial x_1 & \partial B_3 / \partial x_2 & \partial B_3 / \partial x_3 \end{array} \right) = \left( \begin{array}{rrr}
-2.4429 & 9.4865 & -4.5498 \\
4.4430 & 1.4220 & 2.7926 \\
-6.0043 & 0.6362 & 0.8396 \end{array} \right),
\end{equation}
assuming a length scale of 1 and a unit of Gauss. Note that the local electric current (${\bf{J}}$) and the Lorentz force (${\bf{F}}$) is completely determined by $M_{ij}$ as well. The trace of $M_{ij}$ is just $\nabla \cdot {\bf{B}}$ and should vanish. However, because of the linearization (when there might be sub-grid structures) and the computational errors, the zero divergence is not strictly satisfied. We estimate the relative error in calculating $M_{ij}$ to be $|\nabla \cdot {\bf{B}}|\,/\,|\nabla \times {\bf{B}}|=3.2\%$ \citep{xiao2006}. 

The behavior of ${\bf{B}}$ near the singularity is represented by the three eigenvectors ${\bf{v}}_1$, ${\bf{v}}_2$, and ${\bf{v}}_3$ of $M_{ij}$. We find the eigenvectors and their corresponding eigenvalues $\lambda_1$, $\lambda_2$, and $\lambda_3$ to be:
\begin{equation}
\begin{array}{rl}
\lambda_1\,=\,5.4421, & {\bf{v}}_1\,=\,(0.6118, 0.1347, -0.7795)^{\rm{T}}, \\
\lambda_2\,=\,4.4599, & {\bf{v}}_2\,=\,(-0.5130, 0.0380, 0.8575)^{\rm{T}}, \\
\lambda_3\,=\,-10.0833, & {\bf{v}}_3\,=\,(0.7870, -0.4148, 0.4568)^{\rm{T}}.
\end{array}
\end{equation}
In this case, all three eigenvalues are real with one negative and two positives. The eigenvector ${\bf{v}}_3$ with the sole negative eigenvalue $\lambda_3$ determines the initial direction of the ``spine'' field line. The other two eigenvectors ${\bf{v}}_1$ and ${\bf{v}}_2$ define the ``fan'' plane; whereas the linear combination of them gives the initial directions of the ``fan'' field lines. The fan field lines define the separatrix (fan) surface, which separates different domains of magnetic flux.

In practice, field lines traced slightly away from the null in the fan plane tend to flow along the separatrix surface. It is interesting that ${\bf{v}}_1 \cdot {\bf{v}}_2 = -0.9771$, i.e. they are almost 170$^\circ$ with respect to each other. As a result, the traced field lines rapidly converge into two groups, one connecting to N1, the other N2 (Figure~\ref{f:topo}(b)), forming a cusp structure that looks almost two-dimensional. Further analysis classifies this null point as a positive (fan field lines going outward) non-potential null, with current components parallel to the spine and perpendicular to it \citep{parnell1996}.

We note that in some frames in the time series, multiple null points appear in adjacent computational cells near the modeled loop apexes. Both positive and negative nulls exist in the sample, although the morphology of the closed loops remains similar. Such behavior may be related to the uncertainties in modeling and data. More work is needed to evaluate the effect of errors on the inferred topology.







\end{CJK} 

\end{document}